\documentclass[aps,prl,twocolumn,groupedaddress,nofootinbib,showpacs]{revtex4-1}

\usepackage{lipsum}
\usepackage[utf8]{inputenc}
\usepackage{color}
\usepackage{graphicx}
\usepackage{amsmath}
\usepackage{amssymb}
\usepackage{amsfonts}
\usepackage{bm}% bold math
\usepackage{bbm}

\newcommand{\Fpara}{{F_\shortparallel}}

\newcommand{\lpara}{{L}}
\newcommand{\lperp}{{H}}
\newcommand{\dinf}{D_\infty}
\newcommand{\dpara}{D_\parallel}
\newcommand{\bff}{{\bf f}}
\newcommand{\bfF}{{\bf F}}

\newcommand{\bfq}{{\bf q}}
\newcommand{\bfr}{{\bf r}}
\newcommand{\bfrz}{{\bf r}_0}

\newcommand{\bfv}{{\bf v}}

\newcommand{\avg}[1]{{\left\langle#1\right\rangle}}

\begin{document}

\title{Diffusion under confinement: hydrodynamic finite-size effects in simulation}

\author{Pauline Simonnin$^{1,2}$}
\author{Benoit Noetinger$^2$}
\author{Carlos Nieto-Draghi$^2$}
\author{Virginie Marry$^1$}
\author{Benjamin Rotenberg$^1$}
\affiliation{$^1$ \small Sorbonne Universit\'es, UPMC Univ Paris 06, CNRS,
Laboratoire PHENIX, Case 51, 4 Place Jussieu, F-75005 Paris, France}
\affiliation{$^2$IFP Energies Nouvelles, 1 \& 4 avenue de Bois-Pr\'eau, 92852 Rueil-Malmaison}

\date{\today}

\begin{abstract}
We investigate finite-size effects on diffusion in confined fluids
using molecular dynamics simulations and hydrodynamic calculations.
Specifically, we consider a Lennard-Jones fluid in slit pores without 
slip at the interface and show that the 
use of periodic boundary conditions in the directions along the surfaces
results in dramatic finite-size effects, in addition to that of the 
physically relevant confining length.
As in the simulation of bulk fluids, these effects arise from 
spurious hydrodynamic interactions between periodic images 
and from the constraint of total momentum conservation.
We derive analytical expressions for the correction to the diffusion
coefficient in the limits of both elongated
and flat systems, which are in excellent agreement with the
molecular simulation results except for the narrowest pores, where
the discreteness of the fluid particles starts to play a role.
The present work implies that the diffusion coefficients for wide nanopores
computed using elongated boxes suffer from finite-size artifacts which had
not been previously appreciated. In addition, our analytical expression
provides the correction to be applied to the  
simulation results for finite (possibly small) systems.
It applies not only to molecular but also to all mesoscopic hydrodynamic simulations,
including Lattice-Boltzmann, Multi-Particle Collision Dynamics
or Dissipative Particle Dynamics,
which are often used to investigate confined soft matter involving
colloidal particles and polymers. 
\end{abstract}


\maketitle

The dynamics of fluids can be dramatically modified under confinement
down to the molecular scale in nanotubes~\cite{striolo_mechanism_2006,falk_molecular_2010} 
or nanopores~\cite{falk_subcontinuum_2015}, due to the
discreteness of matter and to the interfacial fluid-solid
interactions. Even in larger pores, wider than tens of molecular
sizes which are typical of nanofluidic
devices~\cite{mathwig_electrical_2012,siria_giant_2013}
and for which continuum hydrodynamics hold~\cite{bocquet_nanofluidics_2010},
confining walls influence the dynamics
of the fluids and solutes~\cite{hagen_algebraic_1997,huang_effect_2015}. 
In particular, the diffusion coefficient of particles along a wall is 
generally reduced due to the friction at the interface. 

Quantitatively, the solution of the Stokes equation
in a slit pore involves a series of contributions
of hydrodynamic images situated inside the solid walls. 
It predicts a decrease in the diffusion coefficient 
along the surface with respect to the bulk value, 
of leading order $\sigma/d$, with $\sigma$ the diameter of
the particle and $d$ the distance to the
surface~\cite{faxen_widerstand_1922,blake_fundamental_1974,swan_particle_2010}. 
After averaging over the hydrodynamic slab width 
$\lperp$, this results in a decrease governed by~\cite{SuppMat}:
\begin{align}
\label{eq:dlparainf}
\dpara(\lperp,\infty) 
&\approx\dinf\left[ 1 + \frac{9}{16}\frac{\sigma}{\lperp}
\ln\left(\frac{\sigma}{2\lperp} \right)
\right]
\; ,
\end{align}
where $\infty$ refers to infinite lateral dimensions of the slit
on the left-hand side and to the bulk fluid on the right-hand side.
Using mode-coupling theory, Bocquet and Barrat obtained a similar scaling,   
in good agreement with molecular dynamics (MD) simulations, and emphasized
its origin as the suppression of long-wavelength modes due to
confinement~\cite{bocquet_diffusive_1995}. Slippage at the interface
changes the dependence of the diffusion coefficient with distance
to the surface~\cite{lauga_brownian_2005} and may result in some cases
in an average diffusion coefficient larger than the bulk
value~\cite{saugey_diffusion_2005}.

MD simulations have generally confirmed the decrease in the diffusion
coefficient near a variety of model and realistic 
boundaries~\cite{liu_calculation_2004,marry_structure_2008,sendner_interfacial_2009,
botan_hydrodynamics_2011,wei_molecular_2012,hoang_grand_2012,siboulet_water_2013,
von_hansen_anomalous_2013}
as well as the importance of molecular details in the first adsorbed fluid
layers. However, the use of periodic boundary conditions (PBC) in such
simulations introduces finite-size effects on the diffusion coefficient,
never appreciated under confinement to date, 
due to the distortion of hydrodynamic flows and the spurious hydrodynamic
interaction between periodic images. For bulk fluids in a cubic simulation box,
the correction to the diffusion coefficient
reads~\cite{dunweg_molecular_1993,fushiki_system_2003,
yeh_system-size_2004,tazi_diffusion_2012}
$D(L)= \dinf - \xi k_{B}T/6 \pi \eta L$, with $L$ the box size, $k_B$
Boltzmann's constant, $T$ the temperature, $\eta$ the viscosity
and $\xi\approx2.837$. Recently, the effect of the box shape has also been
considered for bulk fluids: The components of the diffusion
tensor in anisotropic boxes may in some cases be larger than $\dinf$
and diverge with system size in the limit of highly elongated 
boxes~\cite{rozmanov_transport_2012,kikugawa_effect_2015}. 
These features can also be explained from 
hydrodynamics~\cite{botan_diffusion_2015,kikugawa_hydrodynamic_2015,
vogele_divergent_2016}.

Surprisingly, despite the ever growing importance of simulations 
to characterize the dynamics of confined fluids, such finite-size effects
have not been investigated under confinement.
Here we show that the diffusion of confined fluids is not only
affected by the confining distance but is also influenced within simulations
by finite-size effects due to the use of PBC
in the directions parallel to the surfaces. We demonstrate this fact
using MD simulations of a simple fluid confined inside a slit pore. We show
that the diffusion coefficient is generally larger than the value
for the unconfined fluid, by a factor which is in fact significant for
typical box shapes, thereby illustrating the limitations of previous simulation
results. Using continuum hydrodynamics, we further obtain the
scaling with system size in the limits of elongated and flat systems.
As in the case of bulk fluids, the present analysis opens the way to systematic
extrapolation to the limit of infinite systems (for a fixed confining distance).

\begin{figure}[h]
\centering
\includegraphics[width=8.5cm]{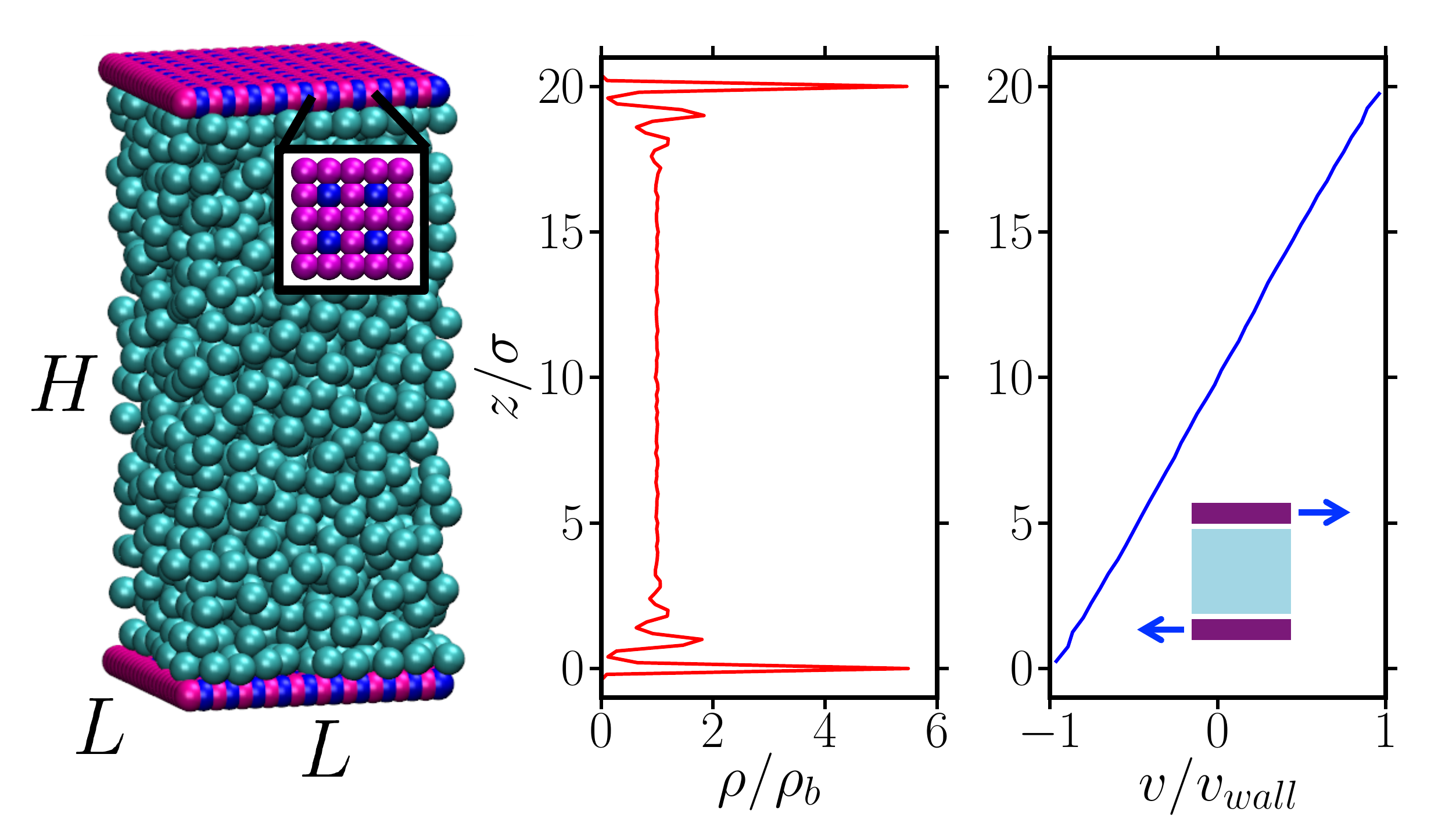}
\caption{The simulated systems consists of a LJ fluid 
between rigid LJ walls. The wall atoms are placed
on a square lattice, with every fourth atom interacting
three times more strongly with the fluid than the others.
In addition to the effect of the confining distance, we investigate
the influence of periodic boundary conditions by varying the size 
$L_x=L_y=\lpara$ of the
simulation box in the directions along the surfaces.
$L_z=\lperp$ indicates the distance between the first adsorbed fluid layers
on both surfaces (the distance between the walls is $\lperp+2\sigma$),
as shown in the central part of the figure for $\lperp=20\sigma$,
which also correspond to the shear planes for such surfaces without
slip at the walls, as demonstrated by the velocity profile under shear.
We consider both elongated ($\lperp>\lpara$, as shown) 
and flat ($\lperp<\lpara$) systems.
}
\label{fig:system}
\end{figure}

We simulate a Lennard-Jones (LJ) fluid under the same conditions as in
previous work illustrating finite-size effects in bulk
fluids~\cite{yeh_system-size_2004,botan_diffusion_2015}, namely
a reduced density $\rho^*=\rho\sigma^3=0.7$ and reduced temperature
$T^*=k_{B}T/\epsilon=2.75$ with $\sigma$ and $\epsilon$ the
LJ diameter and energy, respectively.
The fluid is confined between rigid planar 
surfaces consisting of a square lattice with a spacing of 1$\sigma$.
The fluid-surface interactions are characterized by a diameter $\sigma_{FS}=\sigma$
and an energy $\epsilon_{FS}$ equal to $\epsilon$ for 3/4 of the wall atoms
and $3\epsilon$ for the remaining 1/4 (see the square lattice in
Figure~\ref{fig:system}). 
This pattern ensures the absence of slippage at the wall,
as demonstrated by the velocity profile under shear (with a wall velocity
of $v_{wall}=\pm0.5$ LJ units) in Figure~\ref{fig:system}.
The distance between the LJ walls is $\lperp+2\sigma$, 
with $\lperp$ the distance between the shear planes on both surfaces,
while the size of the simulation box $\lpara$ is the same in the other two directions.
In order to assess the finite-size effects due to both the PBC
and the confinement, we consider systems with $\lpara$ between 6 and
80~$\sigma$ and $\lperp$ ranging from 10 to 160~$\sigma$, corresponding
to particle numbers from 587 to 30720.

All MD simulations are performed using the LAMMPS simulation
package~\cite{LAMMPS}, using a time step of $10^{-3}t^*$, 
with $t^*=\sigma\sqrt{m/\epsilon}$ 
and a cut-off distance $2.5\sigma$ to compute the LJ interactions.
The systems are first equilibrated
in the $NVT$ ensemble during $200t^*$, using a Nos\'e-Hoover thermostat
with a time constant of~$t^*$. After equilibration, trajectories in the $NVE$ ensemble
are produced for $4\,10^{4}t^*$ up to $1.6\,10^{5}t^*$ depending on system size,
from which diffusion coefficients parallel to the surfaces are computed from the
slope of the mean-square displacement (MSD) in the time range
$5\,10^{3}~-~1.5\,10^{4}t^*$ ($4-8\,10^{5}t^*$ for $\lperp\ge80\sigma$
with $\lpara=6\sigma$). 
For each system, reported results correspond to averages and standard errors over 
16 independent runs.

\begin{figure}[h]
\centering
\includegraphics[width=9cm]{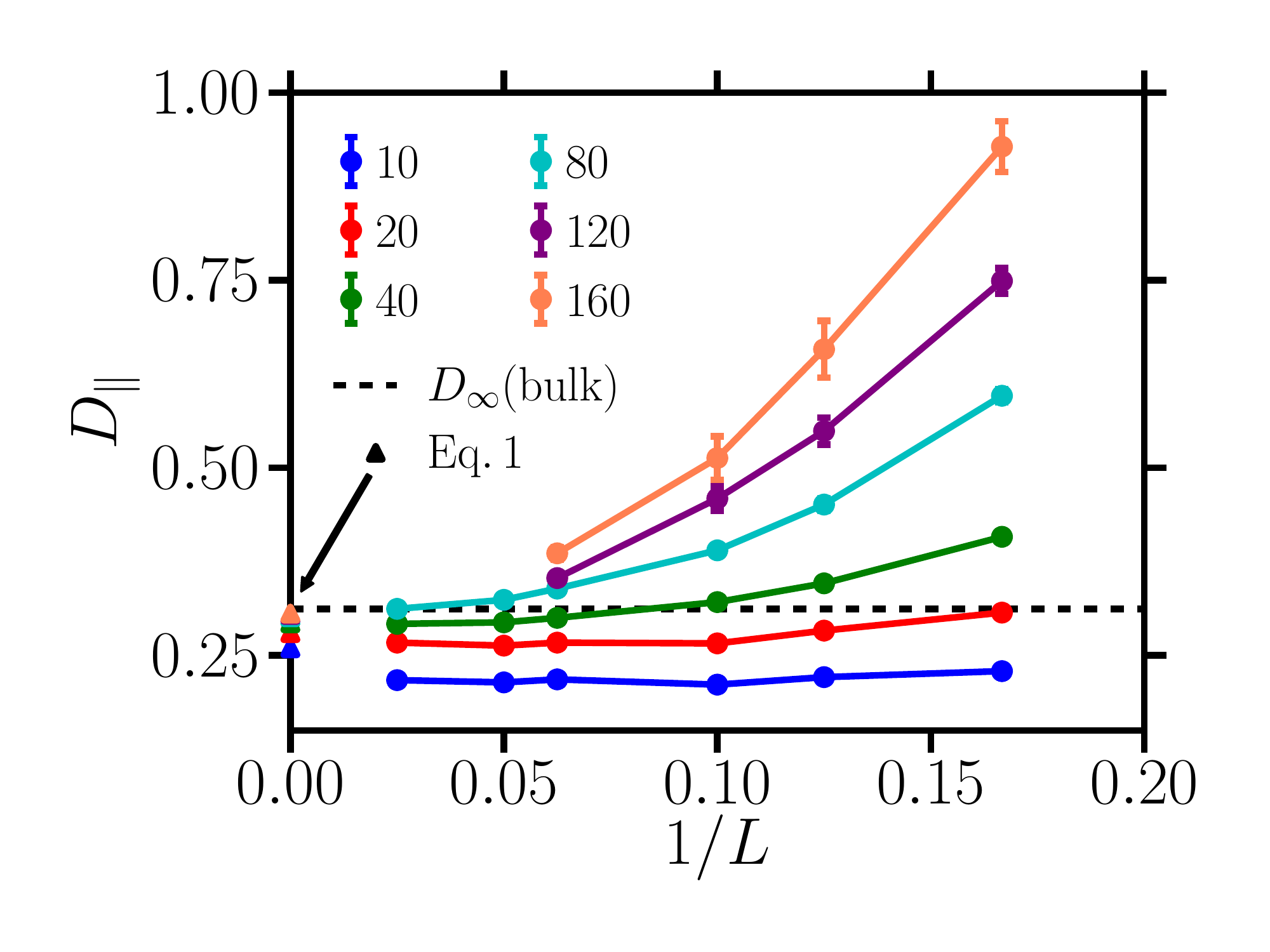}
\caption{Diffusion coefficient along the surface $\dpara$
as a function of the inverse box size parallel to the walls,
$1/\lpara$, for the various confining distances $\lperp$ (all in LJ units).
The horizontal dashed line indicates the extrapolated value for
the bulk fluid in an infinite 
cubic box, $D_{\infty}$, while the values predicted by
hydrodynamics for $\lpara\to\infty$, Eq.~\ref{eq:dlparainf},
are indicated by triangles on the $y$ axis. 
}
\label{fig:DvsinvL}
\end{figure}

Figure~\ref{fig:DvsinvL} reports the diffusion coefficient parallel to the
surfaces $\dpara$ as a function of $1/\lpara$ (by analogy with the
scaling for a cubic box of bulk fluid) for the various confining
distances $\lperp$. This figure also shows the results for an infinite
bulk system, equal to $D_{\infty}=0.312\pm0.005$ LJ units in the present
case~\cite{botan_diffusion_2015}, as well as the prediction of
Eq.~\ref{eq:dlparainf} for the limit $\lpara\to\infty$ under confinement.
We first note that the diffusion coefficient increases with $\lperp$,
as expected, and that the results for large $\lpara$ are consistent with 
Eq.~\ref{eq:dlparainf} (except for the narrowest pore).
However, there is also a dramatic influence of the periodicity along the
surface which increases with pore width $\lperp$. For the smaller $\lpara$,
the diffusion coefficient is larger than $D_{\infty}$ for an infinite unconfined
fluid. For wide thin pores (large $\lperp$, small $\lpara$), it
can be several times larger than $D_{\infty}$.

\begin{figure}[h]
\centering
\includegraphics[width=9cm]{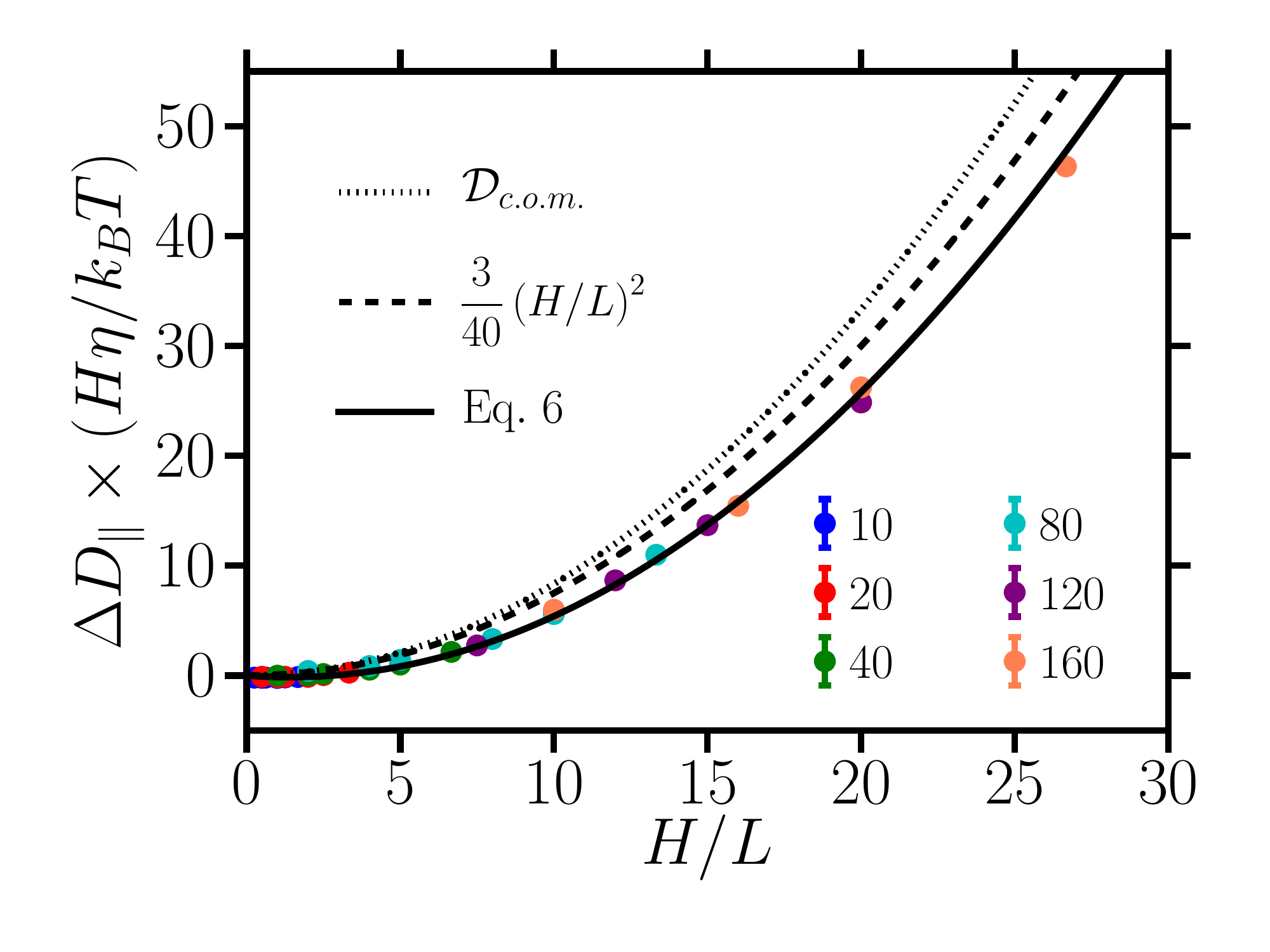}
\caption{Finite size correction to the diffusion coefficient along the surface,
$\Delta\dpara$, with respect to Eq.~\ref{eq:dlparainf} for an infinite
slit pore of width $\lperp$, as a function of $\lperp/\lpara$ with $\lpara$
the size of the periodic system in the directions along the walls.
The dotted line corresponds to Eq.~\ref{eq:dcom}, \textit{i.e.} the
hydrodynamic fluctuations of the center-of-mass;
the dashed line indicates the prediction obtained by treating the 
sum over periodic images (Eq.~\ref{eq:deltamui})
as an integral, while the solid line is the
the full result of Eq.~\ref{eq:delong} which corrects for the 
spurious self-interaction introduced by this assumption.
}
\label{fig:dscaling}
\end{figure}

For elongated boxes ($\lperp>\lpara$), the results of Figure~\ref{fig:DvsinvL}
seem to suggest a correction to the diffusion coefficient proportional to $\lperp/\lpara^2$.
Such a scaling in this regime has been predicted on the basis of hydrodynamic
arguments by Detcheverry and Bocquet, who computed
the enhancement of diffusion in nanometric pipes due to the thermal
fluctuations of the center of mass of the fluid~\cite{detcheverry_thermal_2012,
detcheverry_thermal_2013}. This increase:
\begin{align}
\label{eq:deltad}
\Delta \dpara= \dpara(\lperp,\lpara)-\dpara(\lperp,\infty) 
\;,
\end{align} 
with $\dpara(\lperp,\infty)$ given by Eq.~\ref{eq:dlparainf},
is related to the total fluid-wall friction coefficient $\lambda$ via the Einstein
relation $\Delta \dpara=k_BT/\lambda$. The friction coefficient is then derived
from the total force on the walls exerted by the fluid with typical velocity $v$
as $F\sim\lambda v$, which is also proportional to the fluid-wall contact area 
$2\lpara^2$ and the viscous stress at the boundary $\sim\eta v/\lperp$. 
Therefore, the increase in the diffusion coefficient scales as 
$\Delta \dpara \sim \frac{k_BT}{\eta}\frac{\lperp}{\lpara^2}$.
More precisely, they obtained the following result
for the present slit geometry~:
\begin{align}
\label{eq:dcom}
\Delta \dpara &= \mathcal{D}_{c.o.m.} =  \frac{1}{12}\frac{k_BT}{\eta}\frac{\lperp}{\lpara^2}
\; ,
\end{align}
with $\mathcal{D}_{c.o.m.}$ the diffusion coefficient for the center of mass
of the fluid.  
Figure~\ref{fig:dscaling} reports the simulation results in a
dimensionless form, namely $\Delta\dpara\lperp\eta/k_BT$ as a function
of $\lperp/\lpara$ (here we use the bulk value of the visosity,
$\eta=1.28$ LJ units~\cite{botan_diffusion_2015}).
The collapse of the simulation results on a master curve confirms the hydrodynamic
origin of the finite-size effects. While Eq.~\ref{eq:dcom} accounts
qualitatively for the observed behavior it is not quantitative.

In order to go beyond this simple scaling argument, we now 
compute the correction to the diffusion coefficient due to both confinement 
and PBC along the surfaces by solving the full hydrodynamic problem.
Following previous studies of bulk
fluids~\cite{dunweg_molecular_1993,yeh_system-size_2004,botan_diffusion_2015},
this is achieved by computing the mobility from the solution of the Stokes
equation for an incompressible fluid, $\eta\nabla^2\bfv -\nabla p + \bff = 0$,
with $\bfv$ and $p$ the velocity and pressure fields.
In bulk fluids, the force density $\bff={\bf F}\left[ \delta(\bfr)-\frac{1}{V}\right]$
with ${\bf F}$ a force, $\delta$ the Dirac distribution and $V$ the volume of the system,
includes both a perturbation at $\bfr=0$ and a compensating ``background
force'' which enforces the constraint of vanishing total force on the system.
This compensating force provides in fact the
main contribution to the system-size dependence of the diffusion coefficient.

In the present case, the broken translational invariance in the direction of 
confinement renders the problem more difficult. The mobility tensor 
defined by $\bfv(\bfr)=\mathbb{T}(\bfr,\bfrz)\cdot\bfF(\bfrz)$,
provides the flow at $\bfr=(x,y,z)$ induced by a force $\bfF$ applied at
$\bfrz=(x_0,y_0,z_0)$ and is related to the diffusion tensor as 
$\mathbb{T}=\mathbb{D}/k_BT$.  The expression of the mobility tensor $\mathbb{T}^\infty$
for the case of an infinite slit pore ($\lpara\to\infty$) with non-slip
boundary  conditions at the walls ($\bfv=0$ for $z=0$ and $\lperp$), previously obtained 
by Liron and Mochon~\cite{liron_stokes_1975} 
or Swan and Brady~\cite{swan_particle_2010}, is derived in a convenient form for
the present work in the Supplementary Material~\cite{SuppMat}, which also
provide the details of the following calculations.
It depends on the relative position along the wall
$(x-x_0,y-y_0)$ and on both $z$ and $z_0$.
We note that contrary to the bulk case, the mobility can be obtained
in the presence of solid walls even without compensating background.
The effect of the latter, which remains necessary to enforce the constraint
of total momentum conservation, can be introduced separately by subtracting
the average mobility over the pore volume.
The average correction to the diffusion tensor then reads
$\Delta\mathbb{D} = k_BT\left[ \avg{\delta\mathbb{T}^{i}}-\avg{\delta\mathbb{T}^{b}}\right]$ 
where the first term corresponds to the effect of periodic images
and the second to that of the compensating background.

The effect of periodic images $(m\lpara,n\lpara)$ along the surfaces
can then be expressed in real space as a correction to the mobility
$\delta\mathbb{T}^{i}(x,y,z)=\sum_{(m,n)\ne(0,0)} \mathbb{T}^\infty( x- m\lpara, y -n\lpara, z, z_0=z)$,
with $z_0=z$ since the images are located in the same plane.
We consider here the mobility averaged over the whole pore,
which from translational invariance along the surfaces simplifies to:
\begin{align}
\label{eq:deltamui}
\avg{\delta\mathbb{T}^i} %=\frac{1}{V}\int{\rm d}\bfr\delta\mathbb{T}
= \frac{1}{\lperp}\int_0^\lperp {\rm d}z \hspace{-0.2cm}
\sum_{(m,n)\ne(0,0)} \mathbb{T}^\infty(-m\lpara,-n\lpara, z, z) 
\;.
\end{align} 
The average contribution of the background force, summed over
all periodic images, can be written as an integral over all space:
\begin{align}
\label{eq:deltamub}
\avg{\delta\mathbb{T}^{b}} 
= \frac{1}{\lperp^2}\iint_0^\lperp {\rm d}z{\rm d}z_0
\iint_{-\infty}^{\infty} \frac{{\rm d}x_0{\rm d}y_0}{L^2}
\mathbb{T}^\infty(-x_0,-y_0, z, z_0) 
\;.
\end{align} 
Taking advantage of the symmetry of the system, the component
along the surface is computed from
${\rm T}_\parallel=\frac{1}{2}\left(\mathbb{T}_{xx}+\mathbb{T}_{yy}\right)$.
In addition, the integral over $x_0$ and $y_0$ in $\avg{\delta\mathbb{T}^{b}}$ 
is conveniently computed as the value for $\bfq=0$ of the 2D Fourier transform
$\tilde{{\rm T}}^\infty_\parallel(\bfq, z, z_0)$. 
We show in the Supplementary Material~\cite{SuppMat} that the contribution of the
background is equal to $\avg{\delta{\rm T}_\parallel^{b}}=\lperp/24\eta\lpara^2$.

For elongated systems ($\lperp>\lpara$), the discrete sum over images
in Eq.~\ref{eq:deltamui} 
can be estimated by the corresponding integral,
which is equal to $7\lperp/60\eta\lpara^2$, after removing the
term corresponding to $(m,n)=(0,0)$, which is given by
$-3\ln(1+\sqrt{2})/4\pi\eta\lpara$ (see~\cite{SuppMat} for both demonstrations). 
Subtracting the background, we finally
obtain the correction to the diffusion coefficient:
\begin{align}
\label{eq:delong}
\Delta \dpara(\lperp>\lpara) &= \frac{k_BT}{\eta} \left[ \frac{3}{40}\frac{\lperp}{\lpara^2}
-\frac{3\ln(1+\sqrt{2})}{4\pi\lpara} \right]
\, .
\end{align}
The first term is consistent with the scaling argument of Detcheverry and
Bocquet (see Eq.~\ref{eq:dcom}) and the curvature is only 10\% smaller.
As shown in Figure~\ref{fig:dscaling}, it is closer to the simulation results,
but the second $\mathcal{O}(1/\lpara)$ term, which corrects
for the spurious self-interaction introduced upon replacing the
discrete sum by an integral, is necessary to describe the
simulation results quantitatively. The agreement of Eq.~\ref{eq:delong}
with the latter is excellent.
This confirms the hydrodynamic origin
of the observed finite-size effects due to the PBC, in addition
to the effect of confinement correctly described by Eq.~\ref{eq:dlparainf}
for sufficiently wide pores.

In the opposite regime of flat simulation boxes ($\lperp<\lpara$), 
the mobility ${\rm T}^\infty$ decays exponentially due to
the screening of hydrodynamic interactions by the walls~\cite{SuppMat}.
Therefore the sum (Eq.~\ref{eq:deltamui}) becomes negligible compared to
the effect of the background (Eq.~\ref{eq:deltamub}), so that: 
\begin{align}
\label{eq:dflat}
\Delta \dpara(\lperp<\lpara) &\approx 
-\frac{1}{24}\frac{k_BT}{\eta}\frac{\lperp}{\lpara^2} \, .
\end{align}
This explains the decrease in the diffusion coefficient with $1/\lpara)$
observed for the narrower pores in Figure~\ref{fig:DvsinvL}.
Eq.~\ref{eq:dflat} is in very good agreement with the results
for $\lperp=20\sigma$ (within 5\%) and $40\sigma$ (within 1\%), 
but still overestimates the diffusion
coefficient for $H=10\sigma$ (by $\sim25\%$). 
Since the fluid is perturbed by the surfaces
over 2-3 layers on each wall (see the density profile in
Figure~\ref{fig:system}), it is not surprising that the present continuum 
hydrodynamics calculations, which neglect molecular effects, are not
quantitative in this case.

V\"ogele and Hummer recently obtained
analytical expressions for the correction to the diffusion
tensor for bulk fluids in anisotropic boxes~\cite{vogele_divergent_2016}, 
in good agreement with numerical and molecular simulation 
results~\cite{botan_diffusion_2015}.  For elongated systems, the correction 
for the component corresponding to $\dpara$ scales as
$\frac{k_BT}{\eta}\left[ \lperp/12\lpara^2 - \xi_1/\lpara\right]$,
with $\xi_1\approx0.23$, \textit{i.e.} the same functional
form as Eq.~\ref{eq:delong} but with different numerical factors.
In contrast, for flat systems the scaling goes as
$\frac{k_BT}{\eta}\left[ \ln(\lpara/\lperp)/4\pi\lperp 
- \xi_2/\lperp\right]$, with $\xi_2\approx0.15$
and therefore diverges as $\lpara\to\infty$, which is of course
not the case under confinement. Therefore, despite some
similarities, the present case of confined fluids is fundamentally different
due to the boundary conditions at the solid-liquid interface.
We finally note that, as for bulk fluids, 
the role of the background force enforcing the constraint
of total momentum conservation is essential.
In the bulk, the correction corresponding to the minimum-image cell 
accounts for $\approx85\%$ of the total
correction~\cite{yeh_system-size_2004}. Under confinement,
the total background contribution represents a large
part of the $\mathcal{O}(\lperp/\lpara^2)$ 
term in Eq.~\ref{eq:delong} for elongated systems
(the nearest-image cell corresponds to the
$\mathcal{O}(1/\lpara)$ term)~\cite{SuppMat}
and almost 100\% of the effect for flat systems (Eq.~\ref{eq:dflat}). 

Ideally, the limit $\lpara\to\infty$ should be obtained
from simulations with $\lpara\gg\lperp$ in order to minimize
the finite-size effects due to PBC (see Figure~\ref{fig:DvsinvL}).
The only exception is that of confinement down to the molecular scale,
where such finite-size effects are less dramatic 
(see \textit{e.g.} Ref.~\citenum{holmboe_molecular_2014})
due to the discreteness of the fluid and the predominance of interfacial
features.
However, in practice typical boxes are rather elongated or cubic 
($L\leq H$) than flat, because the latter regime is computationally 
more expensive. 
The present work suggests that it is possible to minimize the finite-size
effects in the other limit of elongated boxes by choosing an aspect ratio
of $H/L=10\ln(1+\sqrt{2})/\pi\approx2.8$ for which $\Delta \dpara$
cancels (see Eq.~\ref{eq:delong}).
It further demonstrates the hydrodynamic origin of these effects on the
diffusion coefficient, as for bulk fluids, and offers with Eq.~\ref{eq:delong}
an estimate of the correction to be applied to the  
simulation results for finite (possibly small) systems.

The present work not only applies to molecular simulations, but also
to all mesoscopic hydrodynamic simulations,
including Lattice-Boltzmann, Multi-Particle Collision Dynamics
or Dissipative Particle Dynamics.
Therefore the effects of PBC should also be taken into account 
in the study of confined soft matter involving colloidal particles and polymers.
The analysis could also be extended to other geometries such as
nanotubes, as well as to slip boundary conditions at the interface,
since slippage is known to have an influence on the diffusion 
coefficient~\cite{lauga_brownian_2005,saugey_diffusion_2005}.
In both cases, the same strategy can be applied using the corresponding
mobility tensor for the limit of an infinite system along the surfaces.

%%%%%%%%%%%%%%%%%%%%%%%%%%%%%%%%%%%%%%%%%%%%%%%%%%%%%%%%%%%%%%%%%%%%%%%%%%%%%%%%%%
%%%%%%%%%%%%%%%%%%%%%%%%%%%%%%%%%%%%%%%%%%%%%%%%%%%%%%%%%%%%%%%%%%%%%%%%%%%%%%%%%%
%%%%%%%%%%%%%%%%%%%%%%%%%%%%%%%%%%%%%%%%%%%%%%%%%%%%%%%%%%%%%%%%%%%%%%%%%%%%%%%%%%
%%%%%%%%%%%%%%%%%%%%%%%%%%%%%%%%%%%%%%%%%%%%%%%%%%%%%%%%%%%%%%%%%%%%%%%%%%%%%%%%%%

%\bibliography{biblio}

%%%%%%%%%%%%%%%%%%%%%%%%%%%%%%%%%%%%%%%%%%%%%%%%%%%%%%%%%%%%%%%%%%%%%%%%%%%%%%%%%%

%\section*{Acknowledgements}

\vskip .5cm
The authors are grateful to Lyd\'eric Bocquet and Jean-Pierre Hansen for useful
discussions.
They acknowledge financial support from IFPEN
and access to computing resources on Curie (TGCC, French National
HPC) via the GENCI project x2016087684 and on MeSU (UPMC).

%%%%%%%%%%%%%%%%%%%%%%%%%%%%%%%%%%%%%%%%%%%%%%%%%%%%%%%%%%%%%%%%%%%%%%%%%%%%%%%%%%%%%%%%%%%%%%%%%

\newpage

\begin{widetext}

\vspace{2cm}
\begin{center}
\textbf{\Large
Supplementary Material
}
\end{center}

\vspace{2cm}

%%%%%%%%%%%%%%%%%%%%%%%%%%%%%%%%%%%%%%%%%%%%%%%%%%%%%%%%%%%%%%%%%%%%%%%%%%%%%%%%%%
%\section{Confined diffusion for an infinite slit pore} 

\noindent
\textbf{\large I. Confined diffusion for an infinite slit pore} 
\vskip .5cm

Here we explain the expression for the hydrodynamic correction to the diffusion
due to the confinement, for an infinite slit pore (Eq.~1) of the main text.
We simply express the results of Saugey \textit{et al.}~\cite{saugey_diffusion_2005} 
with the notations of the present work. The local diffusivity at position $z$
within the pore, defined here by the position of the hydrodynamic shear planes
at $z=\pm H/2$, is assumed to result from the combined frictions
due to the two walls. For the case of non-slip boundary conditions considered
in the present work, this reads:
\begin{align}
D_\parallel(z) &= 
\frac{ D_\infty}{
\displaystyle\frac{1}{1-\displaystyle\frac{9}{16}\frac{\sigma}{2z+\sigma}}
+\displaystyle\frac{1}{1-\displaystyle\frac{9}{16}\frac{\sigma}{2(H-z)+\sigma}}
%+\displaystyle\frac{1}{1-\displaystyle\frac{9}{16}\frac{\sigma}{2H+\sigma-2z}}
-1}
\end{align}
with $D_\infty$ the diffusion coefficient of the unconfined fluid
and $\sigma$ the particle diameter.
Averaging over the pore width, we obtain:
\begin{align}
\frac{\left\langle D_\parallel \right\rangle}{D_\infty} &=
\frac{1}{H}\int_0^H 
\frac{D_\parallel(z)}{D_\infty} {\rm d}z 
\nonumber \\
&= 
1 - \frac{9\sigma}{8H}\sqrt{\frac{7\sigma+16H}{25\sigma+16H}}
\mathrm{Argth}\left( \frac{16H}{\sqrt{(7\sigma+16H)(25\sigma+16H)}} \right)
\end{align}
For wide pores ($H\gg\sigma$) this is approximately given by
$\approx 1 + \displaystyle\frac{9}{16}\frac{\sigma}{H}
\ln\left(\frac{\sigma}{2H} \right)$,
which corresponds to Eq.~1 of the main text. 

%%%%%%%%%%%%%%%%%%%%%%%%%%%%%%%%%%%%%%%%%%%%%%%%%%%%%%%%%%%%%%%%%%%%%%%%%%%%%%%%%%
%\newpage
\vskip 1cm

%\section{Mobility tensor for an infinite slit pore}
\noindent
\textbf{\large II. Mobility tensor for an infinite slit pore} 
\vskip .5cm

In the main text we use the mobility tensor defined by 
$\bfv(\bfr)=\mathbb{T}^\infty(\bfr,\bfrz)\cdot\bfF(\bfrz)$
which provides the fluid velocity $\bf{v}(\bf{r})$
at position $\bfr=(x,y,z)$ induced by a force $\bfF$ applied at
$\bfrz=(x_0,y_0,z_0)$, for an infinite slit pore.
Here we derive its expression.

%%%%%%%%%%%%%%%%%%%%%%%%%%%%%%%%%%%%%%%%%%%%%%%%%%%%%%%%%%%%%%%%%%%%%%%%%%%%%%%%%%
\subsection{Basic hypotheses and equations}\label{Basic}

We consider the flow of a Newtonian fluid at low Reynolds number.  
The motion of the fluid confined between two infinite plates $z=\pm H/2$
parallel to the $xy$ plane is governed by the Stokes equation
$\eta \nabla ^2 \bfv(\bfr)+\bff(\bfr) - \nabla p(\bfr) = {\bf 0}$,
its incompressibility (Eq.~\ref{eq:incomp}) %$\nabla \cdot \bfv(\bfr)=0 $ 
and the non-slip boundary conditions at the wall (Eq.~\ref{eq:boundcond}).
%$\bfv(\bfr) ={\bf 0}$ for $z =\pm H/2$.
%\begin{align}\label{eq:one}
%\eta \nabla ^2 \bfv(\bfr)+\bff(\bfr) - \nabla p(\bfr) &= {\bf 0} \ ,\nonumber \\
%\nabla \cdot \bfv(\bfr) &=0 \ , \nonumber \\
%\bfv(\bfr) &={\bf 0} \ \ {\rm for} \ z =\pm H/2
%\; .
%\end{align}
Here $\eta$ denote the viscosity, while $\bff(\bfr)$ and $p(\bfr)$ denote 
a body force field and the associated pressure.
The mobility tensor $\mathbb{T}^\infty(\bfr,\bfrz)$ is obtained as the solution
of the so-called Stokeslet problem:
\begin{align}
\label{eq:stokes}
\eta \nabla ^2 {\bf{v}(\bf{r})}+\delta(x)\delta(y)\delta(z-z_{0}) \Fpara{\bf{e_x
}} - \nabla p(\bf{r}) & =0 \\
\label{eq:incomp}
\nabla \cdot \bf{v}(\bf{r}) &=0  \\
\label{eq:boundcond}
{\bf{v}(\bf{r})}&={\bf 0}  \ \ {\rm for} \ z =\pm H/2
\end{align}
with a unit horizontal force density ($\Fpara=1$) acting without loss of
generality at $\bfrz=(0, 0, z_{0})$ in the $x$ direction.
This problem was fully solved by Liron and Mochon~\cite{liron_stokes_1975}. 
Their solution involves multiple reflexions at the walls, complex integral representations 
and series expansion techniques, which render the final form difficult to
exploit for the present work. We present here a more direct route,
along the lines of Swan and Brady~\cite{swan_particle_2010}.

%%%%%%%%%%%%%%%%%%%%%%%%%%%%%%%%%%%%%%%%%%%%%%%%%%%%%%%%%%%%%%%%%%%%%%%%%%%%%%%%%%
\subsection{Solution of the Stokeslet problem under confinement}
\label{sec:fullStokeslet}

%Our goal is to compute the  Green's function (Stokeslet) defined by the solution of:
%\begin{eqnarray}\label{eq:twofull}
%\eta \nabla ^2 {\bf{v}(\bf{r})}+\delta(x)\delta(y)\delta(z-z_0)\Fpara{\bf{e_x }} - \nabla p(\bf{r})=0\nonumber \\
%\nabla \cdot \bf{v}(\bf{r})=0 \nonumber \\
%{\bf{v}(\bf{r})}=0, if z =\pm H/2
%\end{eqnarray}
%Here, the body force $\bff_{\shortparallel}$  acts along the x-axis, and it is located at position $x=0, y=0, z=z_0$ of the canal. 

From the Stokes and incompressibility equations (Eqs.~\ref{eq:stokes}
and~\ref{eq:incomp}), the pressure obeys:
\begin{eqnarray}\label{eq:pressure}
\nabla ^2 p(\bfr)=\delta'(x)\delta(y)\delta(z-z_0)\Fpara
\, .
\end{eqnarray}
It is convenient to introduce the partial Fourier transform with respect to $x$ and $y$ defined by:
\begin{eqnarray}\label{eq:fourier1}
h(\bfq,z)= \iint_{-\infty}^{\infty} {\rm d}x{\rm d}y\ h(x,y,z) e^{iq_{x}x+iq_{y}y}
\end{eqnarray}
with ${\bf{q}}=(q_x,q_y)$.
The local velocity may also be decomposed into parallel $\bfv_{\shortparallel}(x,y,z)$ and vertical directions $\bfv_z(x,y,z)$.
Using these notations, the Fourier transforms 
%of $p(\bf{r}), {\bf{v}(\bf{r})}$ denoted respectively by 
$p(\bfq, z)$, $\bfv_{\shortparallel}(\bfq, z)$, and $\bfv_z(\bfq, z)$ are
solutions of:
\begin{align}
\label{eq:FTfullequation}
-q^2 p(\bfq, z)+\frac{\partial ^2  p(\bfq, z)}{\partial z^2}
& = iq_x\delta(z-z_0)\Fpara 
\\
\eta q^2 \bfv_{\shortparallel}(\bfq, z) 
-\eta \frac{{\partial ^2  \bfv}_{\shortparallel}(\bfq, z)}{\partial z^2}
&=\Fpara\delta(z-z_0){\bf e}_x - i\bfq p(\bfq, z) 
\\
\eta q^2 {v_{z}}(\bfq, z) -\eta \frac{\partial ^2 v_z(\bfq, z)}{\partial z^2}
&=\frac{\partial p(\bfq, z)}{\partial z}
\\ 
i\bfq\cdot \bfv_{\shortparallel}(\bfq, z) +\frac{\partial v_z (\bfq,z)}{\partial z}&=0
\label{eq:FTincompress}
\\
\bfv(\bfq, z)&=0  \ \ {\rm for} \ z =\pm H/2
\end{align}
These second order differential equations can be solved using standard methods,
yielding: 
\begin{align}\label{Eq:solfull1}
 \begin{cases}
%\begin{matrix}
p({\bf{q}}, z) &=  A \cosh(qz)+B \sinh(qz)+\frac{iq_x}{2q} \exp(q|z-z_0|)\Fpara\\ 
-\eta v_x({\bf{q}}, z) &=   A^x \cosh(qz)+B^x \sinh (qz)-\frac{iq_x}{2q}z A \sinh (qz)-\frac{iq_x}{2q}z B \cosh(qz) \\
 &\hskip .5cm +\frac{q_{x}^2}{4q^2}|z-z_0|\exp(q|z-z_0|)\Fpara
+\frac{1}{2q}\exp(q|z-z_0|)\Fpara- \frac{q_{x}^2}{4q^3}\exp(q|z-z_0|)\Fpara\\ 
-\eta v_y({\bf{q}}, z)& =  A^y \cosh (qz)+B^y \sinh (qz)-\frac{iq_y}{2q}z A \sinh (qz)-\frac{iq_y}{2q}z B \cosh (qz)\\
&\hskip .5cm +\frac{q_{x}q_{y}}{4q^2}|z-z_0|\exp(q|z-z_0|)\Fpara-\frac{q_{x}q_{y}}{4q^3}\exp(q|z-z_0|)\Fpara  \\ 
-\eta v_z({\bf{q}}, z)& =  A^z \sinh (qz)+B^z \cosh (qz)-\frac{1}{2}A z \cosh (qz)-\frac{1}{2}B z \cosh (qz) \\
&\hskip .5cm -\frac{iq_{x}(z-z_0)}{4q}\exp(q|z-z_0|)\Fpara
%\end{matrix} 
 \end{cases}
\end{align}
where $q=||\bfq||$ and where all the prefactors (to be determined in the
following) depend on ${z_0}$. In the following
this dependence is dropped for the sake of clarity, but it needs to be included when 
performing averages over the pore in the next sections.
One must manipulate quantities such as $(z-z_0)$ and  $|z-z_0|$ with caution, 
especially when computing derivatives of $\exp(q|z-z_0|)$ with respect to $z$.
The eight constants $A, B$, $A^x, B^x$, $A^y, B^y$ and $A^z, B^z$ are determined
from the boundary conditions and the divergence-free condition. 
Differentiating the last equation with respect to $z$, we obtain: 
\begin{align}
-\eta\frac{\partial v_z({\bf{q}}, z)}{\partial z} &=qA^z \cosh (qz)+qB^z \sinh (qz)
-\frac{1}{2}A\cosh (qz)-\frac{1}{2}B \sinh (qz)
\nonumber \\
& \hskip .5cm -\frac{1}{2}A qz \sinh (qz)-\frac{1}{2}B qz \cosh (qz) 
\nonumber \\
& \hskip .5cm -\frac{iq_{x}}{4q}\exp(q|z-z_0|)\Fpara-\frac{iq_{x}|z-z_0|}{4}\exp(q|z-z_0|)\Fpara
\end{align}
Inserting this expression together with $v_x$ and $v_y$ into the incompressibility 
condition Eq.~\ref{eq:FTincompress},
%$i\bfq\cdot {\bf{v_{\shortparallel}}}({\bf{q}}, z) +\frac{\partial v_z ({\bf{q}}, z)}{\partial z}=0$, 
one obtains the following simple conditions:
\begin{align}
\begin{cases}
%\begin{matrix}
\cosh (qz)(iq_xA^x+iq_yA^y+q A^z -\frac{1}{2}A)&=0 \\
\sinh (qz)(iq_xB^x+iq_yB^y+q B^z -\frac{1}{2}B)&=0 
%\end{matrix} 
 \end{cases}
\end{align}
%so that:
%\begin{align}\label{Eq:divbis}
%A&= 2(iq_xA^x+iq_yA^y+q A^z)\\
%B&= 2(iq_xB^x+iq_yB^y+q B^z )
%\end{align}
%or equivalently:
Since this holds for every position $z$, the terms in parentheses must
vanish. These conditions can be written in compact form as:
\begin{align}
\label{eq:ABscal}
\begin{cases}
%\begin{matrix}
A=2 i{\bf Q}\cdot{\bf A} \\
B=2 i{\bf Q}\cdot{\bf B}
%\end{matrix} 
 \end{cases}
\end{align}
where we have introduced the following vectors:
\begin{equation}
{\bf A} = \left(
\begin{array}{c}  
A^x \\ A^y \\A^z  
\end{array} 
\right) \; ; \;
{\bf B} = \left(
\begin{array}{c}  
B^x \\ B^y \\B^z  
\end{array} 
\right)
\; ; \;
{\bf Q} = \left(
\begin{array}{c}  
q_x \\ q_y \\-iq  
\end{array} 
\right)
\end{equation}
Inserting these results in Eq~\ref{Eq:solfull1} and using the non-slip
boundary conditions on $z=\pm H/2$ provide the remaining 6 equations that permit
to determine the unknowns $A$, ${\bf A}$, $B$ and ${\bf B}$. 
In order to proceed as simply as possible,
we rewrite the velocity field under the following form:
\begin{eqnarray}\label{Eq:solAB}
 \begin{cases}
\begin{matrix} 
-\eta v_x({\bf{q}}, z)&=  & A^x \cosh (qz)+B^x \sinh (qz)-\frac{iq_x}{2q}z A \sinh (qz)-\frac{iq_x}{2q}z B \cosh (qz) -\eta v^{0}_x(z)\\ 
-\eta v_y({\bf{q}}, z)& = & A^y\cosh (qz)+B^y \sinh (qz)-\frac{iq_y}{2q}z A \sinh (qz)-\frac{iq_y}{2q}z B \cosh (qz)-\eta v^{0}_y(z) \\ 
-\eta v_z({\bf{q}}, z)& = & A^z \sinh (qz)+B^z \cosh (qz)-\frac{1}{2}A z \cosh (qz)-\frac{1}{2}B z \cosh (qz) -\eta v^{0}_z(z)
\end{matrix} 
 \end{cases}
\end{eqnarray}
with 
\begin{align}
\eta \bfv^{0}(z) &=
\eta \left(
\begin{array}{c}  
v^{0}_x(z)\\ v^{0}_y(z) \\ v^{0}_z(z)
\end{array} 
\right) \; 
%\nonumber\\
%&
= -\;
 \left(
\begin{array}{c}  
     \frac{q_{x}^2}{4q^2}|z-z_0| +\frac{1}{2q}-\frac{q_{x}^2}{4q^3} 
  \\ \frac{q_{x}q_{y}}{4q^2}|z-z_0|-\frac{q_{x}q_{y}}{4q^3}  
  \\-\frac{iq_{x}(z-z_0)}{4q} 
%     \frac{q_{x}^2}{4q^2}|z-z_0|e^{q|z-z_0|} +\frac{1}{2q}e^{q|z-z_0|}-\frac{q_{x}^2}{4q^3}e^{q|z-z_0|} 
%  \\ \frac{q_{x}q_{y}}{4q^2}|z-z_0|e^{q|z-z_0|}-\frac{q_{x}q_{y}}{4q^3}e^{q|z-z_0|}  
%  \\-\frac{iq_{x}(z-z_0)}{4q}e^{q|z-z_0|} 
%     \frac{q_{x}^2}{4q^2}|z-z_0|\exp(q|z-z_0|) +\frac{1}{2q}\exp(q|z-z_0|)-\frac{q_{x}^2}{4q^3}\exp(q|z-z_0|) 
%  \\ \frac{q_{x}q_{y}}{4q^2}|z-z_0|\exp(q|z-z_0|)-\frac{q_{x}q_{y}}{4q^3}\exp(q|z-z_0|)  
%  \\-\frac{iq_{x}(z-z_0)}{4q}\exp(q|z-z_0|) 
\end{array} 
\right) e^{q|z-z_0|} \Fpara
\end{align}
%\begin{equation}
%\bfv^{0}(z) = \left(
%\begin{array}{c}  
%v^{0}_x(z)\\ v^{0}_y(z)\\v^{0}_z(z) 
%\end{array} 
%\right) 
%\end{equation}
By introducing:
\begin{eqnarray}
{\bf V}^{(0)+} = \frac{1}{2}\left[\bfv^{0}(H/2)+\bfv^{0}(-H/2)\right]\\
{\bf V}^{(0)-} = \frac{1}{2}\left[\bfv^{0}(H/2)-\bfv^{0}(-H/2)\right]
\end{eqnarray}
and exploiting the parity of hyperbolic functions,
the six non-slip conditions may be rewritten as:
\begin{eqnarray}
 \begin{cases}
\begin{matrix} 
0&= & A^x \cosh(qH/2)-\frac{iq_x}{2q}\frac{H}{2} A \sinh(qH/2)  -\eta V^{(0)+}_x\\ 
0&= & B^x \sinh(qH/2)-\frac{iq_x}{2q}\frac{H}{2} B \cosh(qH/2) -\eta V^{(0)-}_x \\ 
0&= & A^y \cosh(qH/2)-\frac{iq_y}{2q}\frac{H}{2} A \sinh(qH/2)  -\eta V^{(0)+}_y\\ 
0&= & B^y \sinh(qH/2)-\frac{iq_y}{2q}\frac{H}{2} B \cosh(qH/2) -\eta V^{(0)-}_y \\ 
0&= & A^z \sinh(qH/2)-\frac{A}{2}\frac{H}{2} \cosh(qH/2)  -\eta V^{(0)-}_z\\ 
0&= & B^z \cosh(qH/2)-\frac{B}{2}\frac{H}{2} \sinh(qH/2) -\eta V^{(0)+}_z \\ 
\end{matrix} 
 \end{cases}
\end{eqnarray}
%The quantities $A^x, B^x$, $A^y, B^y$ and $A^z, B^z$ may be expressed with A and B as:
From these relations we express ${\bf A}$ and ${\bf B}$ as a function of $A$ and $B$:
\begin{align}
\label{Eq:solvectorA}
{\bf A} &= \left(
\begin{array}{c}  
     \frac{1}{\cosh(qH/2)}[\eta V^{(0)+}_x+\frac{iq_x}{2q}\frac{H}{2} A \sinh(qH/2)]  
  \\ \frac{1}{\cosh(qH/2)}[\eta V^{(0)+}_y+\frac{iq_y}{2q}\frac{H}{2} A \sinh(qH/2)] 
  \\ \frac{1}{\sinh(qH/2)}[\eta V^{(0)-}_z+\frac{A}{2}\frac{H}{2} \cosh(qH/2)]  
\end{array} 
\right) 
\\
%\; ; \;
\label{Eq:solvectorB}
{\bf B} &= \left(
\begin{array}{c}  
     \frac{1}{\sinh(qH/2)}[\eta V^{(0)-}_x+\frac{iq_x}{2q}\frac{H}{2} B \cosh(qH/2)]
  \\ \frac{1}{\sinh(qH/2)}[\eta V^{(0)-}_y+\frac{iq_y}{2q}\frac{H}{2} B \cosh(qH/2)]
  \\\frac{1}{\cosh(qH/2)}[\eta V^{(0)+}_z+\frac{B}{2}\frac{H}{2} \sinh(qH/2)]  
\end{array} 
\right)
\end{align}
Finally, inserting these relations in Eq.~\ref{eq:ABscal} 
%and the equalities $A=2i {\bf Q}\cdot{\bf A}$ and $B=2i {\bf Q}\cdot{\bf B}$ provides direct relations that allow to determine both constants A and B:
we obtain the expression of $A$ and $B$ as a function of ${\bf V}^{(0)+}$ and
${\bf V}^{(0)-}$:
\begin{eqnarray}\label{Eq:AB}
A=\frac{2i\eta}{1-\frac{qH}{\sinh(qH)}}\left[\frac{1}{\cosh(qH/2)}\lbrace q_x
V^{(0)+}_x +q_y V^{(0)+}_y \rbrace-\frac{iq}{\sinh(qH/2)}V^{(0)-}_z\right]\nonumber \\
B=\frac{2i\eta}{1+\frac{qH}{\sinh(qH)}}\left[\frac{1}{\sinh(qH/2)}\lbrace q_x
V^{(0)-}_x +q_y V^{(0)-}_y \rbrace-\frac{iq}{\cosh(qH/2)}V^{(0)+}_z\right]
\end{eqnarray}
Inserting Eq~\ref{Eq:AB} in Eqs~\ref{Eq:solvectorA} and~\ref{Eq:solvectorB}, and
substituting into Eq~\ref{Eq:solfull1} provides, 
after straightforward although tedious calculations, 
the desired solution for $v_x({\bf{q}}, z, z_0)$.
%the desired solution for $v_x({\bf{q}}, z)=v_x({\bf{q}}, z, z_0)$.

The $xx$ component of the mobility tensor $\mathbb{T}^\infty(\bfr,\bfrz)$
then follows as the inverse 2D Fourier transform
in the particular case $\Fpara=1$, and the $yy$
component can be obtained similarly by computing $v_y({\bf{q}}, z, z_0)$
under a perturbation along the $y$ axis. The parallel component
is finally ${\rm T}_\parallel=\frac{1}{2}
\left(\mathbb{T}^\infty_{xx}+\mathbb{T}^\infty_{yy}\right)$. 
The computation of the inverse 2D Fourier transform is difficult,
but in fact unnecessary. Indeed, for our purpose we only need to
compute averages of the mobility over the pore width. 
This involves integrals over $z$ and $z_0$, as explained
in the next section, as well as over $x$ and $y$ which are
computed directly from the value of the 2D Fourier transform
for $\bfq={\bf 0}$.

%%%%%%%%%%%%%%%%%%%%%%%%%%%%%%%%%%%%%%%%%%%%%%%%%%%%%%%%%%%%%%%%%%%%%%%%%%%%%%%%%%
\subsection{Computation of the associated averages along $z$}

As explained in the main text, two different vertical averages
must be computed. For the effect of the background
$\avg{\delta\mathbb{T}^{b}}$ the average velocity is taken over $z$ and $z_0$
independently:
\begin{align}
\label{eq:verticalaveragefil}
\hat{f}_{z,z_0}^{xx}(\bfq)&=\frac{1}{H^2} \int_{-H/2}^{H/2}\int_{-H/2}^{H/2} 
{\rm d}z{\rm d}z_0\ v_x(\bfq, z,z_0)
\; .
\end{align}
Similarly the $yy$ component is obtained by applying the Stokeslet
in the $y$ direction and computing $v_y$.
This double integral can be performed using the full solution of 
section~\ref{sec:fullStokeslet}, with the result in tensorial form:
\begin{align}
\label{eq:solfil}
\hat{f}_{z,z_0}(\bfq)
&=\frac{(\mathbbm{1}-\hat\bfq\hat\bfq)}{\eta q^2}\frac{1}{H}
\left[1-\frac{\tanh (qH/2)}{qH/2}\right]
\; ,
\end{align}
with $\mathbbm{1}$ the 2D identity matrix and
$\hat\bfq$ a unit vector in reciprocal space.
%The $yy$ component is obtained by replacing $\hat\bfq_x$ by $\hat\bfq_y$,
%so that the half-trace of this tensor (for the parallel component)
The parallel component, given by the half-trace of this tensor,
is particularly simple: 
\begin{align}
\label{eq:resverticalaveragefil}
\hat{f}_{z,z_0}(q)&=\frac{1}{2\eta q^2H}
\left[1-\frac{\tanh (qH/2)}{qH/2}\right]
\; .
\end{align}
Finally, the integral over $x-x_0$ and $y-y_0$ is obtained as
the value of this 2D Fourier transform for $\bfq={\bf 0}$, namely:
%\begin{align}
%\label{eq:limq=0verticalaveragefilsym0}
$\hat{f}_{z,z_0}(0)=H/24\eta$.
%\; .
%\end{align}
Combined with the $1/L^2$ factor for the average in Eq.~5 of the main text,
this leads to the final result:
\begin{align}
\label{eq:background}
\avg{\delta{\rm T}^{b}_\parallel}&=\frac{1}{24\eta}\frac{H}{L^2} 
\; .
\end{align}

For the effect of periodic images $\avg{\delta\mathbb{T}^i}$,
the average is taken over $z=z_0$, \textit{i.e.}: 
\begin{align}
\label{eq:verticalaveragez0z0}
\hat{f}_{z_0,z_0}({\bf{q}})&=\frac{1}{H} \int_{-H/2}^{H/2} {\rm d}z_0\ v_x({\bf{q}}, z_0,z_0)
\; .
\end{align}
The solution in that case is lengthier. The final result for the half-trace
reads:
\begin{align}
\label{eq:FTzzerozzero}
\hat{f}_{z_0,z_0}(q)&=
\frac{
9 + 12q^2H^2 - 2q^4H^4 - 9\cosh(2qH) - 12q^3H^3 \coth(qH) +  9qH\sinh (2qH)
}{48\eta q^2H\left[ \sinh^2(qH) - q^2H^2 \right]}
\end{align}
Here again the value of the integral over  $x-x_0$ and $y-y_0$ is obtained as
the $\bfq={\bf 0}$ value of this 2D Fourier transform, namely:
\begin{align}
\label{Eq:finalsumimagesq=0}
\hat{f}_{z_0,z_0}(0)=\frac{7H}{60\eta}
\; .
\end{align}

%%%%%%%%%%%%%%%%%%%%%%%%%%%%%%%%%%%%%%%%%%%%%%%%%%%%%%%%%%%%%%%%%%%%%%%%%%%%%%%%%%
%\vskip 1cm
\newpage

%\section{Asymptotic results}
\noindent
\textbf{\large III. Asymptotic results} 
\vskip .5cm

As explained in the main text, the correction to the diffusion coefficient
due to periodic boundary conditions is given by  
$\Delta\dpara = k_BT\left[
\avg{\delta{\rm T}^{i}_\parallel}-\avg{\delta{\rm T}^{b}_\parallel}\right]$,
where we have already computed the background correction 
$\avg{\delta{\rm T}^{b}_\parallel}$
in Eq.~\ref{eq:background}. 
Here we derive asymptotic expressions in the regimes of elongated
and flat systems for the effect of periodic images:
\begin{align}
\label{eq:deltamui}
\avg{\delta{\rm T}^{i}_\parallel} %=\frac{1}{V}\int{\rm d}\bfr\delta\mathbb{T}
= \frac{1}{\lperp}\int_0^\lperp {\rm d}z \hspace{-0.2cm}
\sum_{(m,n)\ne(0,0)} {\rm T}_\parallel^\infty(-m\lpara,-n\lpara, z, z) 
\;.
\end{align} 

%%%%%%%%%%%%%%%%%%%%%%%%%%%%%%%%%%%%%%%%%%%%%%%%%%%%%%%%%%%%%%%%%%%%%%%%%%%%%%%%%%
\subsection{Elongated systems ($H\gg L$)}

For elongated systems, we can approximate the discrete sum in Eq.~\ref{eq:deltamui}
by an integral, provided that we remove the contribution corresponding
to $(m,n)=(0,0)$ which is not included in the sum:
\begin{align}
\sum_{(m,n)\ne(0,0)} {\rm T}_\parallel^\infty(-m\lpara,-n\lpara, z, z) 
&\approx 
\iint_{-\infty}^{\infty} \frac{{\rm d}x_0{\rm d}y_0}{L^2}
{\rm T}_\parallel^\infty(-x_0,-y_0, z, z) 
\nonumber \\
& \hskip .5cm
-\iint_{-L/2}^{L/2} \frac{{\rm d}x_0{\rm d}y_0}{L^2}
{\rm T}_\parallel^\infty(-x_0,-y_0, z, z) 
\;.
\label{eq:discretesum}
\end{align} 
The first integral is computed easily using the results of the previous section. 
From Eq.~\ref{Eq:finalsumimagesq=0}, one readily obtains for the average over $z$:
\begin{align}
\label{eq:sumintegral}
\frac{1}{\lperp}\int_0^\lperp {\rm d}z
\iint_{-\infty}^{\infty} \frac{{\rm d}x_0{\rm d}y_0}{L^2}
{\rm T}_\parallel^\infty(-x_0,-y_0, z, z) 
&= \frac{1}{L^2}\hat{f}_{z_0,z_0}(0) 
= \frac{7}{60\eta}\frac{H}{L^2}
\;.
\end{align} 

The second integral can be rewritten as a convolution between
${\rm T}_\parallel^\infty$ and a rectangular function with value 1
if $(x_0,y_0)\in[-\frac{L}{2},\frac{L}{2}]\times[-\frac{L}{2},\frac{L}{2}]$
and zero otherwise. This convolution product is conveniently
computed in Fourier space, using the result Eq.~\ref{eq:FTzzerozzero}
and the well-known transform of the rectangular function.
The average over $z$ then reads:
\begin{align}
\avg{\delta{\rm T}^{\rm self}_\parallel}
&=\frac{1}{\lperp}\int_0^\lperp {\rm d}z
\iint_{-L/2}^{L/2} \frac{{\rm d}x_0{\rm d}y_0}{L^2}
{\rm T}_\parallel^\infty(-x_0,-y_0, z, z) 
\nonumber\\
&= \frac{1}{4\pi^2\eta}
\iint_{-\infty}^{\infty}{\rm d}q_x{\rm d}q_y\
\hat{f}_{z_0,z_0}(q)\frac{\sin(q_xL/2)}{q_xL/2}\frac{\sin(q_yL/2)}{q_yL/2} 
\;,
\end{align} 
with $\hat{f}_{z_0,z_0}(q)$ given by Eq.~\ref{eq:FTzzerozzero}
and $q=\sqrt{q_x^2+q_y^2}$.
We first make a change of variables, $u_x=q_xL$ and $u_y=q_yL$:
\begin{align}
\label{eq:changevariable}
\avg{\delta{\rm T}^{\rm self}_\parallel}
&=\frac{1}{4\pi^2L^2}
\iint_{-\infty}^{\infty}{\rm d}u_x{\rm d}u_y\
\hat{f}_{z_0,z_0}(\frac{u}{L})\frac{\sin(u_x/2)}{u_x/2}\frac{\sin(u_y/2)}{u_y/2} 
\;,
\end{align} 
with $u=\sqrt{u_x^2+u_y^2}$. Now, the regime of elongated boxes
correponds to $H/L\to\infty$, so that we can approximate
$\hat{f}_{z_0,z_0}(\frac{u}{L})$ by the asymptotic expansion of
$\hat{f}_{z_0,z_0}(q)$ for $q\to\infty$, namely:
\begin{align}
\hat{f}_{z_0,z_0}(q\to\infty)
&\approx \frac{3}{8\eta q}
\;,
\end{align} 
which can be derived from the full expression Eq.~\ref{eq:FTzzerozzero}.
Inserting this approximation into Eq.~\ref{eq:changevariable}, we obtain:
\begin{align}
\label{eq:changevariable}
\avg{\delta{\rm T}^{\rm self}_\parallel}
&\approx\frac{1}{\pi^2L^2}\frac{3L}{8\eta}
\iint_{-\infty}^{\infty}{\rm d}u_x{\rm d}u_y\
\frac{1}{\sqrt{u_x^2+u_y^2}}\frac{\sin(u_x/2)}{u_x}\frac{\sin(u_y/2)}{u_y} 
= \frac{3}{8\pi^2\eta L} \times I
\;.
\end{align} 
The integral $I$ defined by the second equality can be computed analytically
by writing:
\begin{align}
\frac{1}{\sqrt{u_x^2+u_y^2}}
= \frac{1}{\sqrt{\pi}}\int_{-\infty}^{\infty} {\rm d}t\ e^{-t^{2}(u_x^2+u_y^2)}
\;.
\end{align}
We then rewrite: 
\begin{align}
I &= \frac{1}{\sqrt{\pi}}\int_{-\infty}^{\infty} {\rm d}t
\iint_{-\infty}^{\infty}{\rm d}u_x{\rm d}u_y\
e^{-t^{2}(u_x^2+u_y^2)}\frac{\sin(u_x/2)}{u_x}\frac{\sin(u_y/2)}{u_y} 
\nonumber \\
&= \frac{1}{\sqrt{\pi}}\int_{-\infty}^{\infty} {\rm d}t
\left[ \int_{-\infty}^{\infty}{\rm d}u\
e^{-t^{2}u^2}\frac{\sin(u/2)}{u}\right]^2  
\nonumber \\
&= \frac{1}{\sqrt{\pi}}\int_{-\infty}^{\infty} {\rm d}t
\left[ \pi\ \mathrm{erf}\left(\frac{1}{4t}\right) \right]^2  
= 2 \pi \ln( 1+\sqrt{2} )
\;.
\end{align} 
In the last line with have introduced the error function and computed the
remaining integral analytically.
Gathering the result with Eq.~\ref{eq:changevariable}, we obtain
$\avg{\delta{\rm T}^{\rm self}_\parallel} = 3\ln( 1+\sqrt{2} )/4\pi\eta L$,
which, together, with Eqs.~\ref{eq:deltamui}, \ref{eq:discretesum} and~\ref{eq:sumintegral}
provides:
\begin{align}
\avg{\delta{\rm T}^{i}_\parallel} 
&= \frac{7}{60\eta}\frac{H}{L^2} - \frac{3\ln( 1+\sqrt{2} )}{4\pi\eta L}
\;.
\end{align} 
Finally, the complete solution for the correction to the diffusion coefficient
is obtained by substracting the contribution of the background,
Eq.~\ref{eq:background}:
\begin{align}
\Delta\dpara( H>L)  &= k_BT\left[ \avg{\delta{\rm T}^{i}_\parallel}
-\avg{\delta{\rm T}^{b}_\parallel}\right]
= \frac{k_BT}{\eta}\left[ \frac{3}{40}\frac{H}{L^2} - \frac{3\ln( 1+\sqrt{2} )}{4\pi L}
\right]
\;,
\end{align} 
which is Eq.~6 of the main text.

%%%%%%%%%%%%%%%%%%%%%%%%%%%%%%%%%%%%%%%%%%%%%%%%%%%%%%%%%%%%%%%%%%%%%%%%%%%%%%%%%%
\subsection{Flat systems ($L\gg H$)}
Ê
Here we show that the mobility tensor ${\rm T}_\parallel^\infty$
decays exponentially fast with distance in real space, so that the interaction between
periodic images $\avg{\delta{\rm T}^{i}_\parallel}$ is negligible
compared to the effect of the background $-\avg{\delta{\rm T}^{b}_\parallel}$.
To that end, we need to compute the inverse Fourier transform of
$\hat{f}_{z_0,z_0}(q)$:
\begin{align}
\label{eq:FTzzerozzeroreal}
\hat{f}_{z_0,z_0}(r)&=\frac{1}{2\pi}\int_{0}^{\infty} {\rm d}q\ qJ_{0}(qr)\hat{f}_{z_0,z_0}(q)
\end{align}
where $\hat{f}_{z_0,z_0}(q)$ is given by Eq.~\ref{eq:FTzzerozzero}
and $J_0$ is the zeroth-order Bessel function of the first kind.
This calculation is much more involved than the previous ones,
which only required the $q\to0$ and $q\to\infty$ limits of $\hat{f}_{z_0,z_0}(q)$.
Liron and Mochon~\cite{liron_stokes_1975} evaluated such integrals using the Hankel
contour illsutrated in Figure~\ref{fig:contour}.
\begin{figure}[htbp]
\centerline{
\includegraphics[width=9cm]{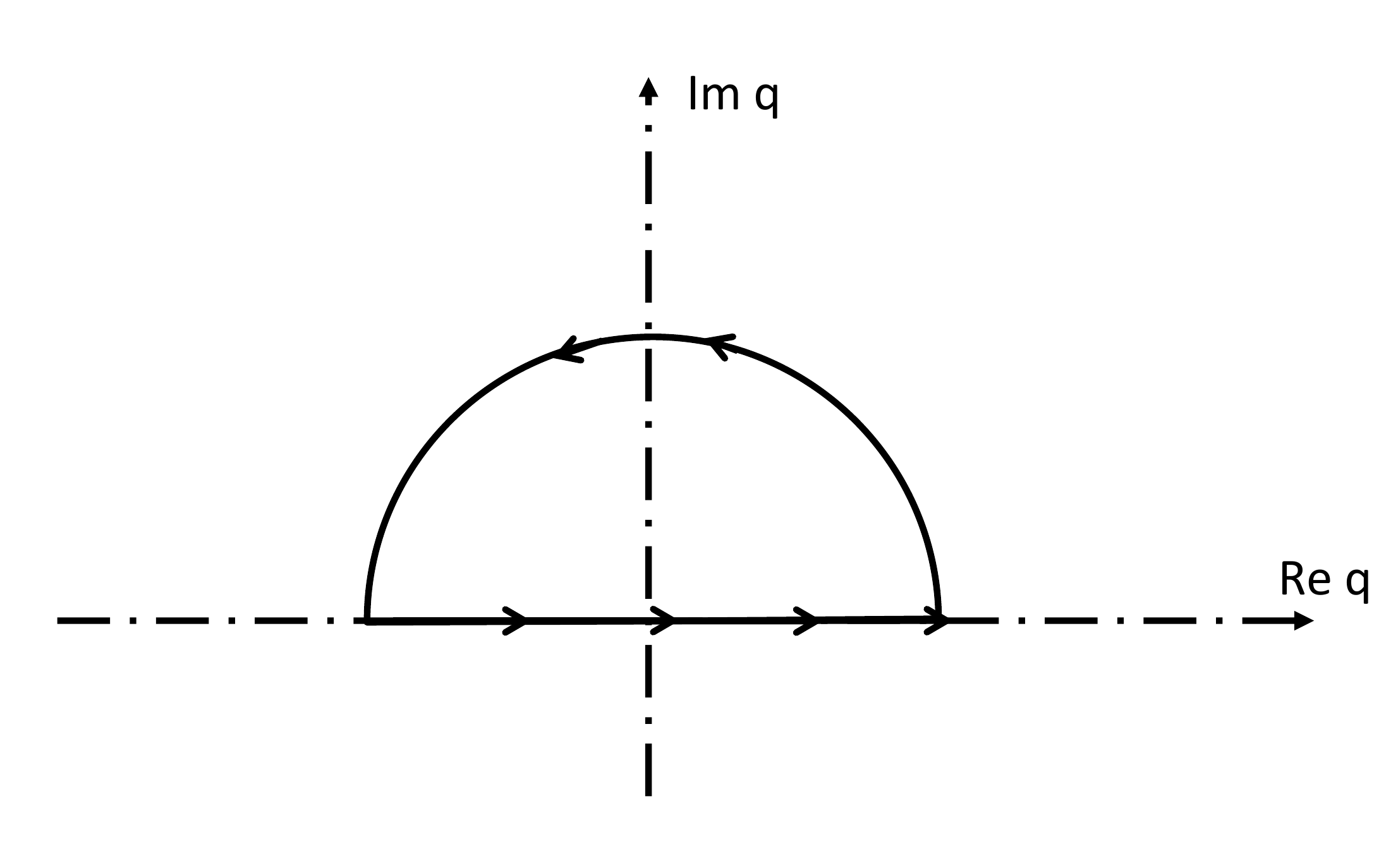}
}
\caption{The integration contour in the complex $q$-plane}
\label{fig:contour}
\end{figure}
Ê
The real-space function is then given by:
\begin{align}
\label{eq:FTzzerozzeroreal}
\hat{f}_{z_0,z_0}(r)&= {\rm Re}\left(\frac{i}{2}
\times\text{sum of residues in upper half plane of } \hat{f}_{z_0,z_0}(q)q H_0^{(1)}(qr) 
\right)
\end{align}
where the function $\hat{f}_{z_0,z_0}(q)$ given by Eq.~\ref{eq:FTzzerozzero}
is now understood as a function of the complex variable $q$
and is itself complex valued. The function $H_0^{(1)}$ 
is a Hankel function given by $H_0^{(1)}(z) =J_0(z)+iY_0(z)$,
with $Y_0$ the zeroth-order Bessel function of the second kind.
The function $\hat{f}_{z_0,z_0}(q)q H_0^{(1)}(qr)$
has no pole at the origin, but an infinite number of poles corresponding to the
$\coth$ function (except at the origin) and additionnal poles denoted
$s_n$ corresponding to the solutions of the transcendental equation
$\sinh(s)^2=s^2$. For large $n$ the asymptotic behavior of these poles is given by:
\begin{align}
\label{eq:asymptpoles}
s_n=x_n+iy_n \simeq \ln (2n+1)\pi +i(n+1/2)\pi.
\end{align}
The net result appears as an absolutely convergent series of functions involving
modified Bessel functions $K_0(n\pi r/H)\propto \displaystyle\sqrt{\frac{H}{r}}\exp (-n\pi r/H)$
coming from the residues arising from the $\coth$ function. An additionnal
series of Bessel functions 
$|H_0^{(1)}(s_n r/H)|\propto
\displaystyle\frac{\pi}{2}\sqrt{\frac{H}{r}}\exp (-y_n r/H)$ comes from the residues
associated with the $s_n$ poles. As in both cases, the imaginary part of the
poles behaves as an arithmetic sequence, the associated series can be
approximated by their first term for large $r/H$.
Overall, the asymptotic behavior for large $r/H$ is
given by:
\begin{align}
\hat{f}_{z_0,z_0}(r)
\sim \sqrt{\frac{H}{r}}\exp (-\pi r/H)
\;,
\end{align}
\textit{i.e.} an exponentially decreasing correction.Ê 
This result can in fact be recovered using equation (49) of Liron and
Mochon~\cite{liron_stokes_1975}, by computing the required trace that suppreses
the long-range contribution, and performing the associatedÊ$(z_0,z_0)$ average
term by term in the resulting series.
The exponential decay in real space implies that
the sum over periodic images is dominated for flat systems
($L\gg H$) by the nearest images, so that this sum also decays exponentially
fast with $L/H$. Overall, confinement screens the hydrodynamic interactions
between the periodic images and the corresponding contribution
$\avg{\delta{\rm T}^{i}_\parallel}$ is negligible
compared to the effect of the background $-\avg{\delta{\rm T}^{b}_\parallel}$.

\end{widetext}

%%%%%%%%%%%%%%%%%%%%%%%%%%%%%%%%%%%%%%%%%%%%%%%%%%%%%%%%%%%%%%%%%%%%%%%%%%%%%%%%%%

\end{document}